# Coverage Control of Silica-Assisted Grown DNA Crystals**

*Junwye Lee, Sunho Kim, Junghoon Kim, Chang-Won Lee, Yonghan Roh,* Sung Ha Park*

The impetus behind the current interest in combining DNA materials to conventional nanotechnologies, such as nanoelectronics,[1,2] biosensors,[3,4] and nanophotonics,[5,6] emanates from an ambition to exploit its remarkable properties.[7] One of these properties is self-assembly which is driven by the thermodynamics of sticky end hybridization and makes structural DNA nanotechnology a prime candidate for bottom-up fabrication schemes in these fields. However, unless self-assembled DNA nanostructures can be fabricated on solid surfaces to at least the degree of accuracy of existing top-down methods, it will be unfeasible to replace it with existing technologies. An intermediate step toward this goal has been to merge the two approaches such that DNA nanostructures are self-assembled onto lithographically patterned substrates. Previous works have been successful at depositing self-assembled DNA nanostructures on patterned substrates[8,9] and controlling the spatial orientations of tailored DNA origami motifs at specifically designated sites,[10,11,12] all of which have used random depositions (or similar methods) of preformed DNA structures onto lithographically patterned substrates. What has been lacking in the literature is a method of precisely controlling the coverage, i.e., the percentage of the surface covered by crystals, of DNA structures on various substrates, especially silica ($SiO_2$), which is crucial if DNA is to be universally employed in electronics. We provide a solution to this problem by introducing a new surface assisted fabrication method, termed the silica-assisted growth (SAG) method, to self-assemble DNA nanostructures on $SiO_2$ surfaces. The novel fabrication technique presented here bears two important distinctions from previous studies. First, direct annealing on the substrates allows for very accurate control of the amount of DNA structures that self-assemble on the substrate, i.e., the coverage. Second, due to electrostatic interactions with the silica surface, structures grown by this method show drastic topological changes leading to previously unreported novel structures.

The pretreatment process of $SiO_2$ substrates and the various DNA structures grown on them are shown in Figure 1. Silanol groups on the $SiO_2$ surface become deprotonated once the substrates are treated in a 10×TAE/$Mg^{2+}$ buffer since the pH of this buffer exceeds the isoelectric point of $SiO_2$.[13] This allows $Mg^{2+}$ ions to bind to the substrate surface which in turn binds the negatively charged DNA backbones (Figure 1a).[10] To demonstrate the SAG method, four different types of DNA nanostructures were prepared. 8 helix tubes (8 HT) and 5 helix ribbons (5 HR) were constructed from single-stranded tiles (SST),[14] while double-crossover (DX) crystals and DX crystals with biotin modifications were fabricated from DX tiles (Figure S1, 2, and 3 in the Supporting Information).[15] The schematic diagrams of the various DNA structures are illustrated in Figure 1b-f and their corresponding atomic force microscopy (AFM) images on $SiO_2$ substrates are shown in Figure 1g-p (Figure 1g-k show structures made from the free solution annealing method deposited onto $SiO_2$ for imaging and Figure 1l-p show structures made using the SAG method where the structures are annealed directly on the substrate, Figure S4 in the Supporting Information).

For the 8 HT, there is a dramatic difference between the structure formations of the free solution annealing method and SAG method. Due to a local minimum in the free energy landscape,[14] monodisperse 8 HT structures on $SiO_2$ fabricated from the free solution annealing method are stable which can be clearly seen in Figure 1g. In this case, the structures have already been formed in the solution before they are deposited onto the substrate. Meanwhile, in the case of SAG, the $Mg^{2+}$ charges bound on the substrate surface interact with the DNA strands to prevent the formation of tubes. Through these interactions, an acute topological change of the structures occurs, allowing SSTs to bind edgewise and remain in a single layer state (Figure 1l) with some of the tiles overlapping along their boundaries (Figure 1l, bright regions). To the extent of the authors' knowledge, this is the first observation of 2D crystals arising from SST motifs. Another type of 1D structure, the 5 helix ribbon, was also successfully fabricated using both the free solution annealing and SAG methods as can be seen in Figure 1h and m, respectively. The substrate acts as a catalyst to reduce the amount of energy needed for DNA structures to form, resulting in large-scale structure formations on the substrates.[16]

In the case of DX crystals, two-tile units of DX monomers were used as building blocks to fabricate periodic arrays. One of the key factors limiting the employment of DNA nanostructures in applications has been the lack of control of the coverage of the various DNA motifs on substrates. For DX crystals grown from the

[*] J. Lee,[+] J. Kim, Prof. S. H. Park
Sungkyunkwan Advanced Institute of Nanotechnology (SAINT) and Department of Physics
Sungkyunkwan University
Suwon 440-746, Korea
E-mail: sunghapark@skku.edu

S. Kim,[+] Prof. Y. Roh
School of Information and Communication Engineering
Sungkyunkwan University
Suwon 440-746, Korea
E-mail: yhroh@skku.edu

Dr. C. -W. Lee
Samsung Advanced Institute of Technology (SAIT)
Yongin 446-712, Korea

[+] These authors contributed equally to this work.

[**] This work was supported by the SAIT-SAINT Research Cooperation Program (2010-2011) funded by Samsung Electronics & the Basic Science Research Program through the National Research Foundation (NRF) of Korea funded by the Ministry of Education, Science and Technology (2010-0013294) to S.H.P and by the National Research Foundation (NRF) of Korea funded by the Korean government (MEST) (No. R01-2008-000-20582-0) to Y.R.

Supporting information for this article is available on the WWW under http://www.angewandte.org or from the author.

free solution annealing method (Figure 1i), the crystals prefer to assemble in the longitudinal direction of the tiles, leaving large sections uncovered by crystals. In stark contrast, due to changes in the aspect ratio of the crystal, crystals grown with the SAG method (Figure 1n) form large domains which possess some mobility on the substrates. Control of the coverage by these domains can be achieved by adjusting the concentration of the DX monomers, [DX] (Figure S5 in the Supporting Information). Furthermore, the functionality of the DX crystals remains unchanged when grown with the SAG method as DX crystals with biotinylated oligonucleotides (DXB) were successfully fabricated (Figure 1e, j, o). As with ordinary DX crystals, a drastic difference in the coverage between the free solution annealing method (Figure 1j) and the SAG method (Figure 1o) was observed. The periodic attachment of the biotin molecules to the DX crystals becomes much more apparent when streptavidin is added, a protein with a high binding affinity for biotin, to the solution (Figure S6 in the Supporting Information). Figure 1f illustrates a schematic diagram of the DXB crystal containing streptavidin proteins (DXB-S), which can be seen in the AFM images as bright dots for the free solution annealing method (Figure 1k) and the SAG method (Figure 1p). In all the experiments concerning DX crystals (DX, DXB, and DXB-S), major topological transformations occurred and exact control of the coverage was possible with the SAG method.

The coverage dependence on the DX monomer concentration for the SAG method was measured by annealing different concentrations of DNA strands (Figure 2).** Figure 2a represents a schematic illustration of the SAG annealing process and Figure 2b-g show AFM images of crystal formations with increasing [DX]. Assembly of DX crystals begin to appear at a threshold concentration of 10 nM (Figure 2c), i.e., the monomer saturation concentration, $m_s$. The actual nucleation of the crystals may begin at a lower concentration than 10 nM, implying that $m_s$ may be lower than 10 nM, but due to accuracy limitations in the experiment (AFM resolution, deviations in the pipette volume, etc.) we have set $m_s$ = 10 nM. As mentioned above, crystals formed on the substrate are topologically different from ordinary free solution DX crystals. The saturation concentration where the $SiO_2$ substrate is completely covered by a monolayer of DX crystals is 20 nM (Figure 2f). As with 5 HR and 8 HT, when structures are grown using SAG, the substrate acts to lower the activity, and thus the free energy, as crystallization of DX crystals occurs slightly below 10 nM instead of typical free solution concentrations of ~50 nM. The coverage dependence on the concentration is shown in Figure 2h and a schematic diagram of the crystal growth is shown in Figure 2i. As [DX] is increased and passes the monomer saturation concentration, $m_s$, the DX tiles start to crystallize on the substrate and continues to grow until a crystal saturation concentration, $c_s$, is reached at which point the substrate is fully covered by monolayers of DX crystals. Thus, by controlling [DX], accurate control of the coverage of DX crystals on the substrate is possible, from 0 to 100%. ***

Another crucial factor in applications of DNA nanostructures is the accurate formation of the structures onto lithographical patterns. Figure 3a shows the patterning process of $SiO_2$. Photoresist (PR) patterns were formed by covering a photoresist layer deposited on a $SiO_2$ substrate with a mask and exposing it to UV light. After development, the substrates were dipped in a octadecyltrichlorosilane (OTS) solution.[17] The PR patterns were removed and the substrate treated in a piranha solution after which the substrate was incubated in a 10×TAE/$Mg^{2+}$ buffer solution for deprotonation. This patterned substrate was then used in the SAG annealing process with the intent that DNA structures would only form on the surfaces of the $SiO_2$ substrate and not on OTS monolayers due to a Coulomb attraction between the DNA strands and $Mg^{2+}$ treated regions of the substrate and the prevention of $Mg^{2+}$ ions binding to the hydrophobic methyl-terminated OTS monolayers (Figure 3a). The binding of $Mg^{2+}$ ions to $SiO_2$ and not to OTS surfaces can be verified by electric force microscopy (EFM). Untreated OTS monolayers have a higher electric potential compared to $SiO_2$ surfaces whereas the situation is reversed once $SiO_2$ is treated with a 10×TAE/$Mg^{2+}$ buffer solution.[18] Successful patterning of DNA monolayers using this method was confirmed through AFM images (Figure 3b-i). The yellow lines indicate the boundaries between the crystals of the DNA monolayers (dark regions) and the OTS monolayers (bright regions). Line and square patterns of DXB monolayers are shown in Figure 3b, c, d and Figure 3e, respectively, where the accuracy of the DNA growth matching the lithographic patterns can be seen. This becomes more conspicuous once streptavidin is added (Figure 3f-i). The bright lines of the crystals in Figure 3g and i are due to DXB crystals in which streptavidin has been attached (DXB-S). The precision of the DNA growth can be observed in the square patterns (Figure 3h and i) where the DXB-S crystals have grown to exactly fit the rounded corners of the squares (Figure S8 in the Supporting Information). Analyses of all the images confirm a very high degree of DNA pattern accuracy.

The range of applications which may benefit from this fabrication scheme seems very broad. $SiO_2$-integrated nanostructures and devices with novel physical, chemical, and biological properties resulting from the incorporation of DNA are now one step closer to becoming a reality. Biologically templated monolayers for biomolecular sensors,[19,20] DNA monolayers modified with cetyltrimethylammonium to form electron blocking layers in solar cells,[21] light-emitting diodes,[22,23] host materials in active waveguide structures in electro-optical devices,[24] and gate dielectrics in field-effect transistors[25,26] are just a few examples. Further works to expand the applicability of surface assisted growth techniques beyond oxide surfaces to metal and polymer surfaces are underway. If successful, these schemes would greatly promote the utility of DNA crystals to almost all types of materials at the micro and nanometer scale.

## Figure legends

*Figure 1.* DNA nanostructure fabrication by silica-assisted growth (SAG). a) Pretreatment of $SiO_2$ substrates in a piranha solution, followed by incubation in a 10×TAE/$Mg^{2+}$ buffer solution and its consequent deprotonation. $Mg^{2+}$ ions readily bind to the deprotonated surface of the substrate. b-f) Schematic diagrams of the various DNA nanostructures made by the free solution annealing and SAG methods; 8 helix tubes (b), 5 helix ribbons (c), double-crossover (DX) crystals (d), double-crossover crystals with biotin (DXB) (e), and streptavidin bound double-crossover (DXB-S) crystals with biotin (f). g-p) AFM images of the structures fabricated by the free solution annealing method (g-k) and the SAG method (l-p). AFM images in the same column as the schematic diagrams of the DNA structures (b-f) correspond to those structures. 8 HT structures, fabricated from an SST motif, in a free solution are 1D crystals (g), but change into 2D polycrystalline structures by using the SAG method (l). Images of DX crystals (n-p) grown by the SAG method show 100% coverage. The insets in (n) and (o) are noise-filtered reconstructed images by fast Fourier transform showing the periodicity of the crystals. The inset in (p) is a magnification of the DXB-S crystal. The scale bars in all the AFM images are 1 µm unless otherwise noted.

*Figure 2.* Coverage control by DX monomer concentrations. a) Schematic diagram illustrating the kinetic process of crystal synthesis on a silica substrate. b-g) The coverage (CR) of silica can be straightforwardly controlled by increasing DX monomer concentrations ([DX]). Noise-filtered fast Fourier transform images showing no signs of crystal periodicity (inset of b) and definite crystal periodicity (insets of f and g). h) A plot of the coverage versus the DX monomer concentration. A coverage of 100% is reached at [DX] = 20 nM and persists till 200 nM, which was the highest concentration tested. i) A schematic plot of the number of monomers versus [DX]. The rate of increase of the number of monomers in the test tube ($\alpha_t$ = 6.0) is greater than the rate of increase of the number of monomers on the substrate ($\alpha_s$ = 0.4) for $m_s \leq [DX] \leq c_s$. This indicates that higher levels of supersaturation in the solution need to be achieved for crystal growth on the substrate as [DX] is increased. The scale bars in all the AFM images are 1 µm unless otherwise noted.

*Figure 3.* Silica-assisted grown DX crystals on patterned silica substrates. a) Process of lithographically patterning $SiO_2$ substrates. The EFM image verifies regular patterns of potential differences between the embossed (OTS layer) and depressed (deprotonated $SiO_2$) features of the substrate. b-e) AFM images of a monolayer of DX crystals grown using SAG on a patterned substrate with gap widths of 2 µm (b), 3 µm (c), 5 µm (d), and a square pattern of 5 µm × 5 µm (e). f-i) AFM images of DXB-S crystals grown using SAG on a patterned substrate with gap widths of 3 µm (f), 5 µm (g), a square pattern of 5 µm × 5 µm with the inset showing a magnified image of the corner of the square detailing the intricate growth of the DXB-S crystal (h), and square islands of DXB-S crystals self-assembled on a patterned silica substrate (i). The scale bars in all the AFM images are 1 µm unless otherwise noted and the yellow lines are guides for the eyes.

## Experimental Section

**Pretreatment of $SiO_2$ substrates.** 300 nm-thick $SiO_2$ layers were thermally grown on p-type silicon substrates. The $SiO_2$ wafers were cleaned by piranha solution for 30 minutes, followed by rinsing with DI-water. The cleaned wafers were cut into 0.5 × 0.5 $cm^2$ pieces with a diamond tip pen and immersed into a microtube filled with 1 ml of 10×TAE/$Mg^{2+}$ buffer solution (400 mM Tris, 10 mM EDTA (pH 8.0), 125 mM magnesium acetate) for 3 hours, followed by rinsing with DI-water.

**DNA Annealing.** Synthetic oligonucleotides, purified by high performance liquid chromatography (HPLC) were purchased from Integrated DNA Technologies (IDT, Coralville, IA). The details can be found on www.idtdna.com. Complexes were formed by mixing a stoichiometric quantity of each strand in physiological buffer, 1×TAE/$Mg^{2+}$. A final concentration of 200 nM is achieved. For annealing, the substrate along with the DNA strands were inserted into AXYGEN-tubes which were then placed in a Styrofoam box with 2L of boiled water and cooled slowly from 95 °C to 20°C over a period of at least 24 hours to facilitate the hybridization process.

**Streptavidin binding to DX-Biotin Nanostructures.** Biotinylated oligos were purchased from www.idtdna.com. Streptavidin was purchased from Rockland Inc. (PA, USA). A 200 nM solution of streptavidin was prepared in DI water. A 1:1 ratio of streptavidin-SAG DXB was prepared by directly pippetting streptavidin solution in the test tube.

**AFM Imaging.** For AFM imaging, a SAG sample was placed on a metal puck using instant glue. 30 µL 1×TAE/$Mg^{2+}$ buffer was then pippetted onto the substrate and another 5-10 µL of 1×TAE/$Mg^{2+}$ buffer was dispensed into the AFM tip (Veeco Inc.). AFM images were obtained by Multimode Nanoscope (Veeco Inc.) in liquid tapping mode.

**Preparation process of $SiO_2$ patterns.** 2 µm-thick photoresist (AZ5214) patterns consisting of arrays of lines and squares were formed on the $SiO_2$ wafers through a photolithography processes. To form an OTS self-assembled monolayer, the wafers were immersed into a solution of OTS (0.1 mM) and hexane for 1 hour. Subsequently, the wafers were immersed and sonicated in acetone to strip photoresist patterns. Residual photoresists were removed and selective hydrophilic patterns formed by immersing OTS/$SiO_2$ patterns into a piranha solution for 3 min and rinsing with DI-water. After 3 minutes of piranha treatment, the OTS regions of the $SiO_2$ substrates remain hydrophobic whereas the $SiO_2$ layer turns hydrophilic.

**EFM Imaging.** For EFM imaging, a $Mg^{2+}$-treated patterned silica substrate was placed on a metal puck using instant glue and blown using a nitrogen gun. EFM images were obtained by SPA400 AFM (Seiko) under a tip voltage of 10 V and tip frequency of 23 kHz.

** In order to achieve a consistent and reliable diagram, three crystal growth conditions, i.e., the substrate size (5 mm × 5 mm), total DNA sample volume (500 µl), and annealing time (~24 hours in 2 L of water in a styrofoam box) were fixed, while different DNA monomer concentrations were used as a control parameter.

*** It should be noted that the crystal growth rate on the substrate, $\alpha_s$, is lower than the rate of the increase of monomers in the free solution of the test tube, $\alpha_t$, by a factor of ~1/15, as [DX] is increased from $m_s$ to $c_s$, shown in Figure 2i (Table S1 and Figure S7 in the Supporting Information). Under normal saturation conditions for $m_s \leq$ [DX] $\leq c_s$, $\alpha_t$ should equal $\alpha_s$, but the fact that $\alpha_t$ is greater than $\alpha_s$ implies that due to the finite size of the substrate, crystals which have already formed on the substrate suppress newly forming crystals from attaching to it, thus making the need to supersaturate the free monomer concentration in the test tube in order to increase the number of crystal formations on the silica substrate.

Figures

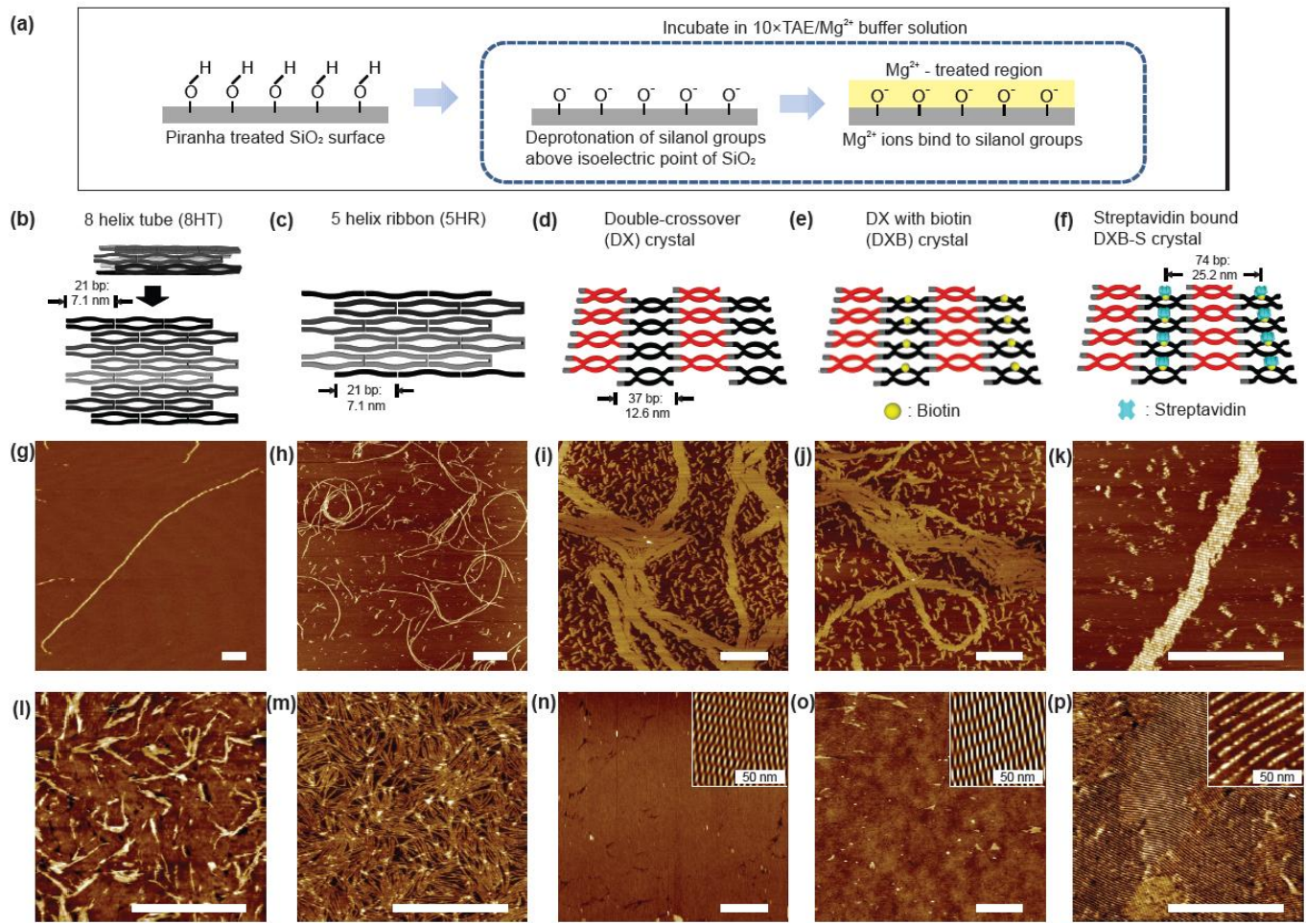

Figure 1

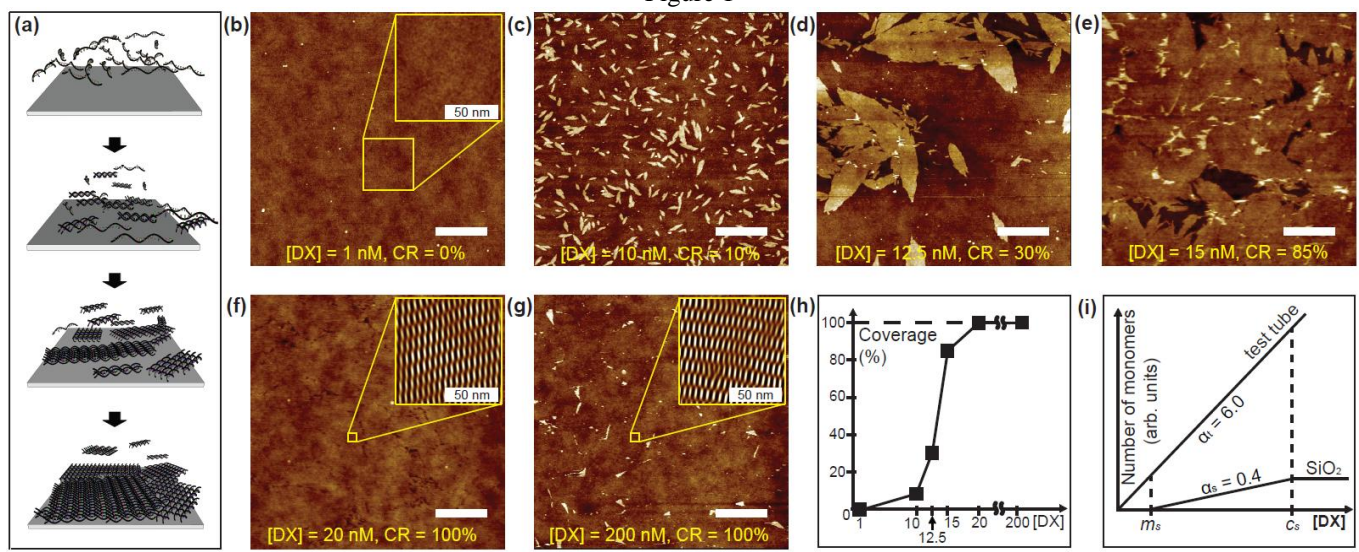

Figure 2

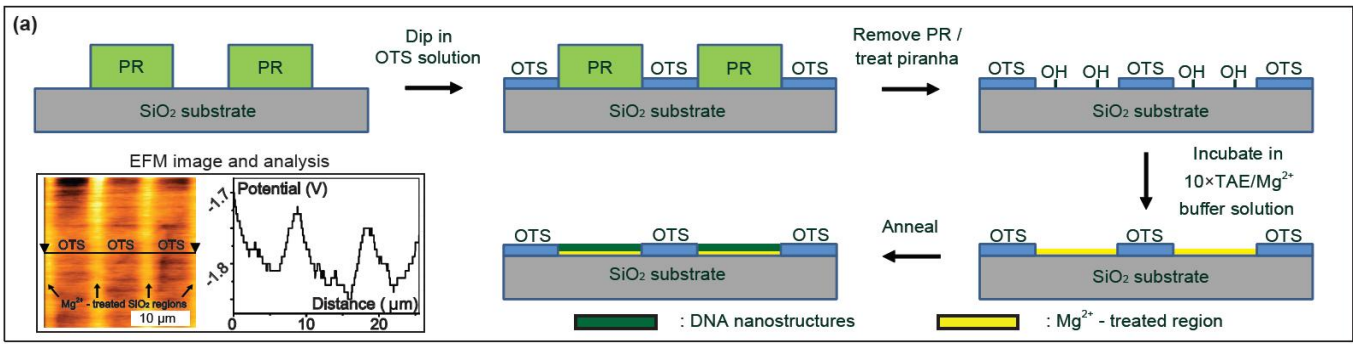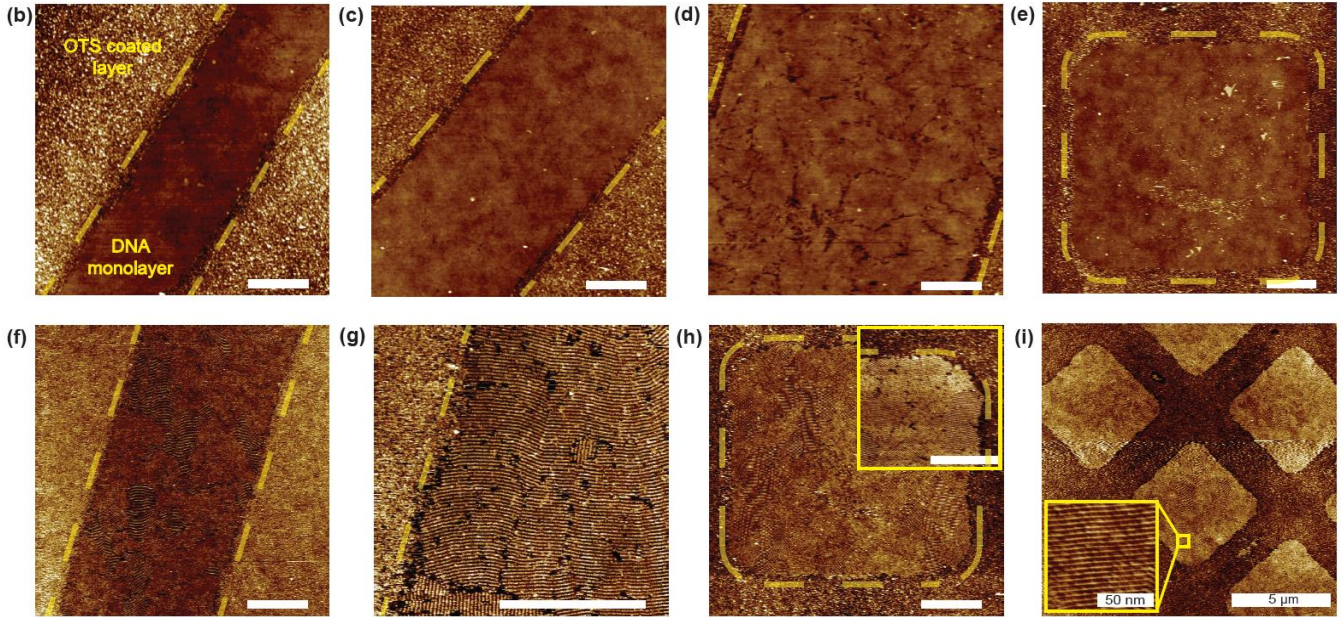

Figure 3